\input{aipcheck}

\documentclass[sort&compress,
    ,final            
  ]
  {aipproc}

\layoutstyle{6x9}

\begin{document}

\title{Anomalous diffusion and quasistationarity in the HMF model }


\author{Alessandro Pluchino}{
  address={Dipartimento di Fisica e Astronomia and INFN, Via S. Sofia 64, 95123 Catania, Italy}
}

\author{Andrea Rapisarda}
{
  address={Dipartimento di Fisica e Astronomia and INFN, Via S. Sofia 64, 95123 Catania, Italy}
}

\begin{abstract}
 We explore  the quasistationary regime 
 of the Hamiltonian Mean Field Model (HMF) showing that at least three different 
 classes of events exist, with a different diffusive behavior and with 
 a relative frequency which depends on the size of the system. 
 Along the same line of a recent work \cite{epl}, 
 these results indicate that one must be very careful in
exchanging time averages with ensemble averages during the non-ergodic 
metastable regime and at the same time they emphasize
the role of finite size effects in the evaluation of the diffusive properties of the system.

\end{abstract}

\maketitle


\section{Introduction}

It is a common practice in statistical physics to exchange time averages with ensemble averages 
since it is  usually assumed that the ergodic hypothesis is in general valid. 
Although the latter is very often verified, it is not always true, expecially for complex systems.  
In a recent paper \cite{epl} we discussed one example where this happens in the
context of the well known Hamiltonian mean field (HMF) model, a paradigmatic long-range 
system whose behavior  has been 
very debated in the last decade \cite{hmf1,hmf2,pluchino,epn,antoniazzi}.
Working in the  same direction, in this paper we want to focus our attention 
on  a controversial topic regarding the anomalous dynamics 
of the HMF model, i.e. its  superdiffusive behavior   observed in the metastable 
regime \cite{epn,adiff1,adiff2,adiff3,adiff4,hmf-dif,yama,bouchet,moyano}. 
In particular, we present new numerical results 
which permit to clearly identify at least three classes of events in that regime, 
showing  a different temperature evolution and  a different diffusive behavior. 
The relative frequency of the three types of events strictly depends on the size of the system. 
These results indicate that one must be very careful in exchanging  time averages with ensemble averages 
 and at the same time  they emphasize the role of finite size effects in the evaluation  
 of the diffusive properties of the system.

\section{Model and Anomalies}

The HMF model describes a system of $N$ fully-coupled classical inertial
XY spins (rotators)
$\stackrel{\vector(1,0){8}}{s_i} = (cos~\theta_i,sin~\theta_i)~,~i=1,...,N,
$with unitary module and mass \cite{hmf0}.
These spins can also be thought as particles rotating
on the  unit circle.
The  Hamiltonian can be written as  
\begin{equation}
\label{hamiltonian}
        H
= \sum_{i=1}^N  {{p_i}^2 \over 2} +
  { 1\over{2N}} \sum_{i,j=1}^N  [1-cos( \theta_i -\theta_j)]~~,
\label{eq.2}
\end{equation}
where ${\theta_i}$ 
is the angle
and $p_i$ the conjugate variable representing the rotational
velocity of spin $i$.
\\
At equilibrium the model can be solved exactly and 
one has  a second order phase transition from a high
temperature paramagnetic  phase to a low temperature
ferromagnetic  one \cite{hmf0}. 
The order parameter of this phase transition is the modulus of
the {\it average magnetization} per spin defined as:
$M = (1 / N) | \sum_{i=1}^N
\stackrel{\vector(1,0){8}}{s_i} | ~~$.
The transition occurs at a critical temperature $T_c=0.5$, which 
 corresponds to a critical energy per particle $U_c = E_c /N =0.75$.
Above $T_c$,  rotators  point towards  different
directions and $M \sim 0$.
Below $T_c$, at variance, they are aligned and trapped 
into  a single cluster, so that  $M \neq0$.
The out-of equilibrium dynamics of the  model is also very interesting. In a range
of energy densities between $U\in[0.5,0.75]$, special  initial conditions called
\textit{water-bag}, with  initial magnetization $M_0=1$ (i.e. with all the spins aligned
and with all the available energy in the kinetic form),
drive the system, after a violent relaxation,
towards metastable Quasistationary States (QSS), where the system remains
trapped for a while before slowly relaxing towards equilibrium. These QSS
are characterized by a lifetime  
which diverges with the system size $N$ \cite{hmf2,pluchino,epn}
and by a temperature $T_{QSS}$ which results to be lower than the canonical equilibrium one.
In the thermodynamic limit, $T_{QSS}$ tends to a limiting value $T_{N\rightarrow\infty}$
which depends on the energy density.
\\
Numerous dynamical anomalies characterize the QSS regime, e.g. 
vanishing Lyapunov exponents, non-Gaussian velocity distributions, slow
decaying velocity correlations, fractal-like phase space structures, aging and
anomalous diffusion \cite{hmf2,pluchino,epn}.
Among them, the diffusive behavior of the  rotators  is one of the most debated.
In order to study diffusion it is customary to consider the mean square displacement
of phases $\sigma^{2}(t)$ defined as
\begin{equation}
\sigma^{2}(t) = \frac{1}{N} \sum_{j=1}^{N}
  [ \theta_{j}(t) - \theta_{j}(0) ]^{2}
  = < [ \theta_{j}(t) - \theta_{j}(0) ]^{2} > ~,
\label{msd2}
\end{equation}
where the symbol $<...>$ represents the average over all the $N$
rotators. Following the one-dimensional generalized Einstein's relation, 
the quantity $\sigma^{2}(t)$, if it is finite, typically scales as $\sigma^{2}(t)\sim t^{\gamma}$: 
the diffusion is normal when $\gamma=1$ (corresponding to the well known law for Brownian
motion) and ballistic for $\gamma=2$ (corresponding to free
 motion). For different values of $\gamma $ the diffusion 
is anomalous and in particular for $\gamma>1$ one has superdiffusion. 
\\
\begin{figure}
\includegraphics[height=.5\textheight]{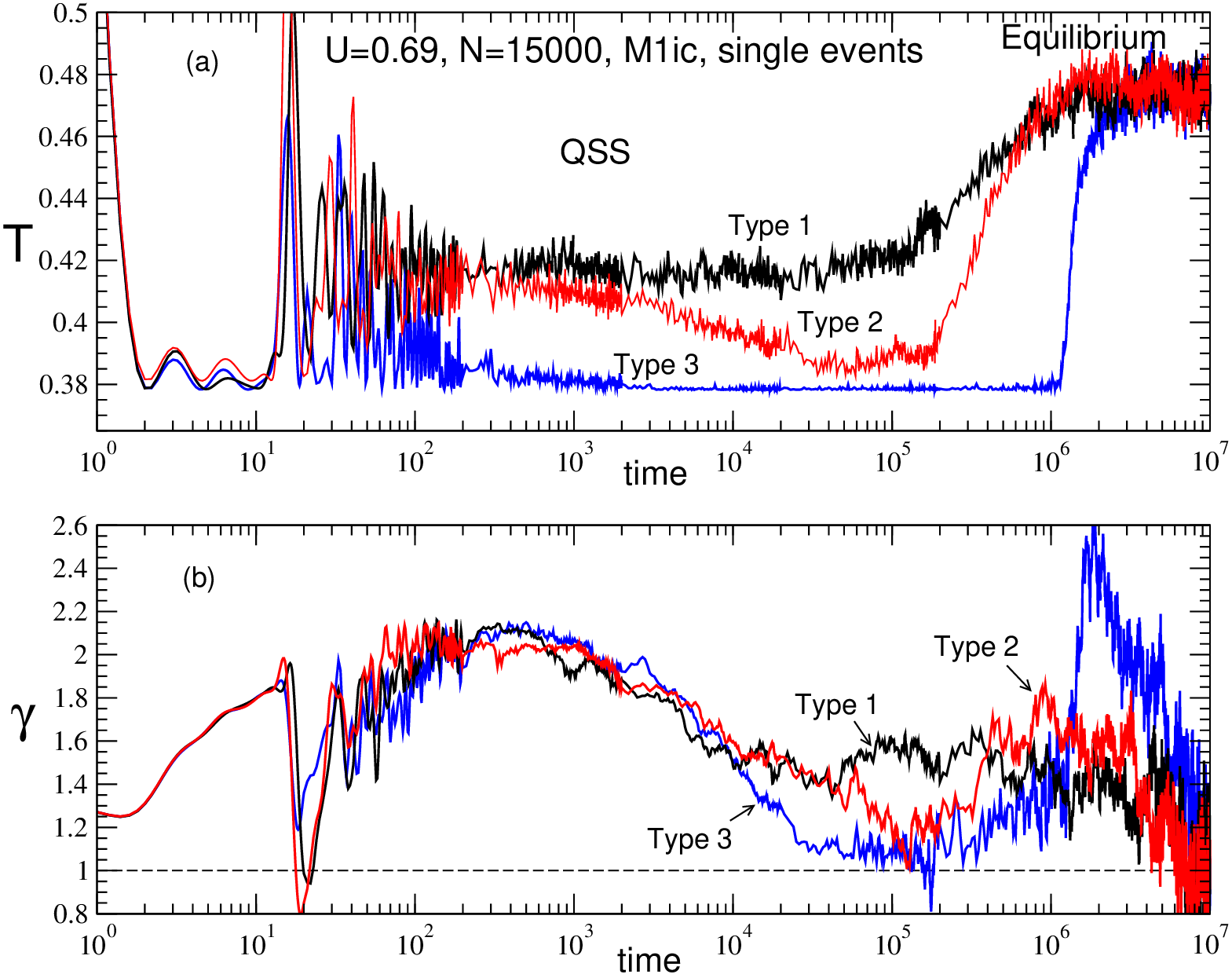}
\caption{(a) Temporal evolution of the temperature $T$ for three representative realizations (events) 
of the $M_0=1$ initial conditions, for a system of $N=15000$ rotators at $U=0.69$. 
(b) Temporal evolution of the diffusive exponent $\gamma$ for the same three events.}
\end{figure}
Superdiffusion has been so far observed in the metastable regime of the HMF model 
for water-bag initial conditions with variable initial  magnetization $M_0$.
 It has been found an exponent $\gamma$ going progressively from $1.4-1.5$ (for $0.4<M_0<1$) to
$1$ for $M_0=0$ \cite{hmf-dif,adiff1,adiff2}.
More recently, a general relationship between
the slow decay of the velocity autocorrelation functions and the superdiffusive behavior, 
based on a theoretical result  by Tsallis and Bukman \cite{tsa-buk},
has been proposed \cite{adiff1,adiff2}. This formula  has been the object of
several controversies \cite{yama,bouchet,ruffo}, also related to the possible application 
of the non-extensive statistical mechanics in this context.
The main objection is that anomalous diffusion in the QSS regime is likely  
only a finite size effect and therefore in  an infinite system one  should recover
$\gamma\sim1$. Furthermore, it has recently  been claimed \cite{ruffo} 
that if in the $QSS$ regime the velocity autocorrelation functions
can be fitted by a $q$-exponential, then the diffusion should be normal,
in apparent contradiction with the results of \cite{adiff1,adiff2,epn}.
\\
In the next section, supported by new numerical simulations, we shall discuss 
these points in detail.

\section{Discussion of numerical results}

In Fig. 1 we consider a system of $N=15000$ rotators at the energy density $U=0.69$.  This energy density 
has been  well studied  in the past being the  value  at which  the anomalies are most evident.
We have adopted  an initial magnetization $M_0=1$. 
In Fig. 1 (a) we plot the temporal evolution of temperature $T$ (calculated 
dynamically as $T(t)=\frac{1}{N}\sum_i {p_i}^2$) for three different realizations (events) 
of the initial conditions. The latter have been  selected among many others as representative of  
three classes
of events  which are observed most frequently. 
Please see \cite{pluchino} for details about the HMF equations of motion and the integration algorithm
adopted. In the   figure,  
it is clearly visible that, after a short violent relaxation stage,
where the system suddenly relaxes from the initial high temperature state, in all the three cases the
system enters into a longstanding metastable regime at a temperature $T_{QSS}$ lower than the canonical equilibrium one
(for $U=0.69$ one has the equilibrium temperature  $T_{eq}\sim0.476$) and only for $t>10^5$ definitively 
relaxes towards equilibration.
But the temperature plateaux appear very different  in the three cases. In fact  
for the \textit{type 1} event,  the temperature oscillates around $T\sim0.41$ and starts to slowly
relax towards equilibrium after $t\sim10^5$; 
for the  \textit{type 2} event,  the temperature stays for a while
around $T\sim0.40$, then  relaxes towards $T\sim0.38$ (\textit{double} plateau) 
and finally, after $t\sim10^5$, slowly reaches the equilibrium value;
for the  \textit{type 3} event,  the  temperature stays around $T\sim0.38$ 
(i.e. the limiting $T_{N\rightarrow\infty}$ for $U=0.69$) up to $t\sim10^6$ and then 
abruptly relaxes to equilibrium.
 The description of these three kinds  of events refers to the case $N=15000$ of Fig.1, but we have found 
 a similar behavior for a wide range of system sizes. So we can identify more generally three different 
 classes of events  
 
 \begin{itemize}

	\item \textit{Type 1} event -  the system shows  a single quasistationary plateau at  temperature $T_{QSS}$ with $T_{N\rightarrow\infty} < T_{QSS} < T_{eq}$ where it remains  for a long time before equilibration;

\item  \textit{Type 2} event - the system passes through two different quasistationary plateaux before equilibration: the first one is similar to that of type 1 event, while the second is at $T_{N\rightarrow\infty}$; 

\item \textit{Type 3} event - the system exhibits again a  single  plateau, but at   temperature $T_{N\rightarrow\infty}$, where it  stays for a longer time with respect to type 1 event   and relaxes then abruptly to  equilibrium.

\end{itemize}
 
  Since in the literature the QSS temperature plateaux have been always
calculated performing averages over many events, such a different behavior of single runs has been
overlooked. Moreover, if one considers  small system sizes,  fluctuations   can hide  such a different behavior, 
which emerges  very clearly only for large  sizes  $N \geq 10000$. 
The difference  among these three  main classes  of events open  new perspectives for  the dynamical anomalies registered 
during the QSS regime, as we discuss immediately below. 
\\
In Fig. 1 (b),  we plot the instantaneous diffusive exponent 
$\gamma(t)$ as a function of time, calculated for the same three events previously discussed
by taking the logarithm of both sides of the Einstein's
generalized relation and differentiating with respect to $ln~t$:
\begin{equation}
  \gamma(t)=\frac{d(ln~\sigma^2)}{d(ln~t)}  ~~~~.
  \label{gamma}
\end{equation}
Note that the time scales of the two panels of Fig. 1 are the same, so that  it is possible to compare
the transitions to the different regimes. It clearly appears that to the three types
of events plotted in the top panel corresponds to a different diffusive behavior in the bottom one.
After a common ballistic regime ($\gamma\sim2$) between $t=10^2-10^3$, in all the three cases $\gamma(t)$ 
starts to decrease; but only for the \textit{type 3} event it quite monotonically reaches 
the value $\gamma\sim1$ indicating normal diffusion (and remains there apart a big peak 
due to the sudden temperature relaxation towards equilibrium). 
In fact, for the \textit{type 1} event $\gamma(t)$ stays around $1.5$ (the same value found in 
\cite{adiff1,adiff2,epn}) up to $t\sim10^6$, then - when the system
equilibrates - definitively relaxes to $\gamma\sim1$. Finally, the \textit{type 2} event  
shows a behavior oscillating between the previous two.
\\
\begin{figure}
\includegraphics[width=0.6\textheight]{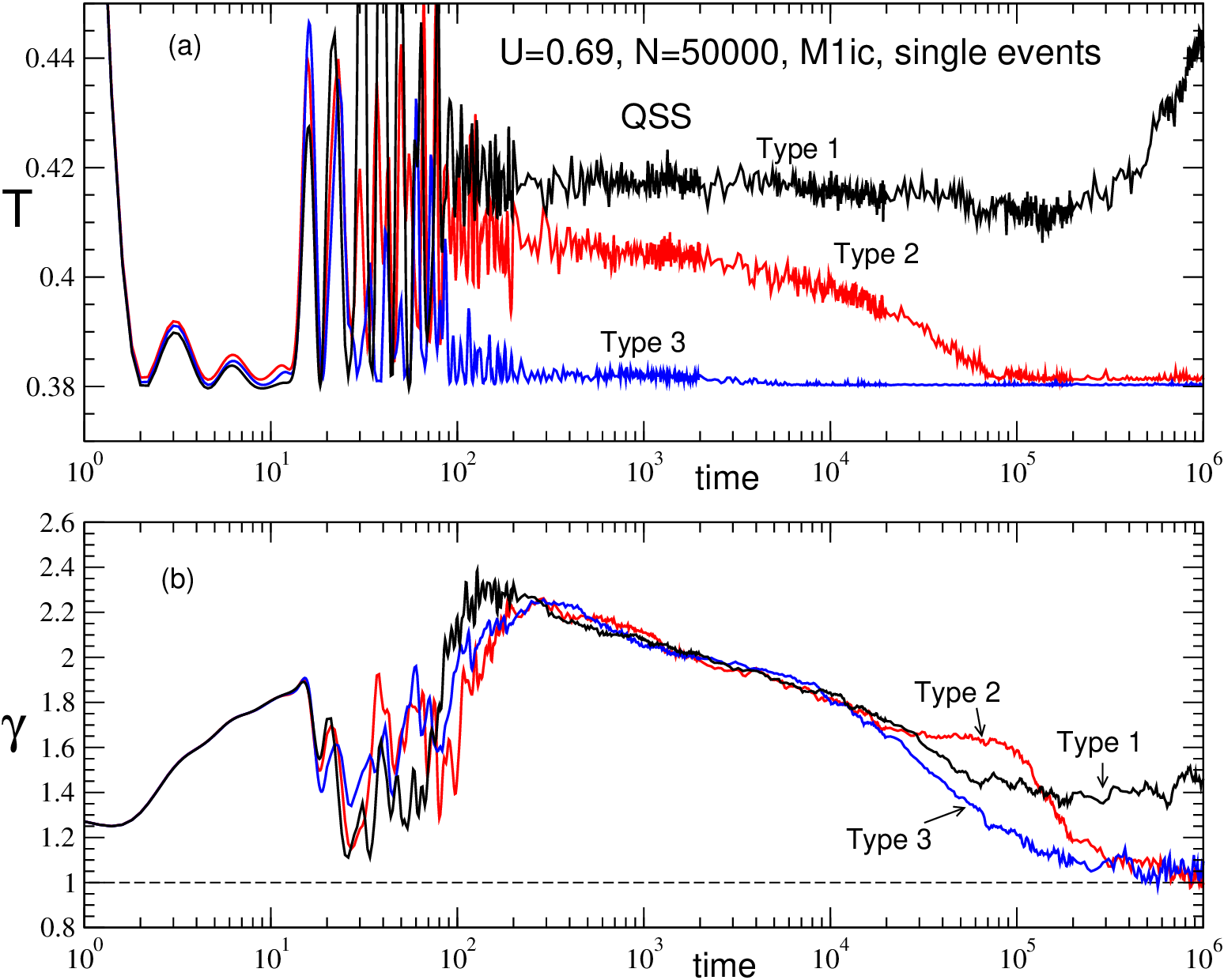}
\caption{(a) Temporal evolution of the temperature $T$ for three representative realizations (events) 
of the $M_0=1$ initial conditions, for a system of $N=50000$ rotators at $U=0.69$. 
(b) Temporal evolution of the diffusive exponent $\gamma$ for the same three events.}
\end{figure}
In Fig. 2 we repeat the same simulations of Fig. 1 but for a larger system of $N=50000$ rotators, in order
to reduce the fluctuations. The results seem to confirm, in a even more evident way, the previous picture: 
again we recover three types of events with a different diffusive behavior, and again only in the 
\textit{type 3} event, whose temperature directly stabilizes in the $T_{N\rightarrow\infty}\sim0.38$ plateau,
the system monotonically reaches the normal diffusion unitary value. On the contrary, in the \textit{type 1} 
event $\gamma(t)$ shows again a plateau around the value $1.5$ for $t>5\cdot10^4$, that persists also during
the relaxation towards equilibrium. Finally, the intermediate \textit{type 2} event shows again a \textit{double}
QSS temperature plateau, much clearer than that shown in the $N=15000$ case: this time
$\gamma(t)$ firstly monotonically decays towards the value $1.65$, where remains for $5\cdot10^4<t<1\cdot10^5$; 
then, when the temperature reaches the limiting value $T_{N\rightarrow\infty}\sim0.38$, it 
slowly relaxes towards $\gamma\sim1$. For this size of the system we do not plot the complete relaxations
to equilibrium for the three events but they approximatively follow the behavior of Fig. 1, with $\gamma(t)$
that reaches in all the cases the value $1$ when the temperature reaches its equilibrium value.
\\
\begin{figure}
\includegraphics[width=0.5\textheight]{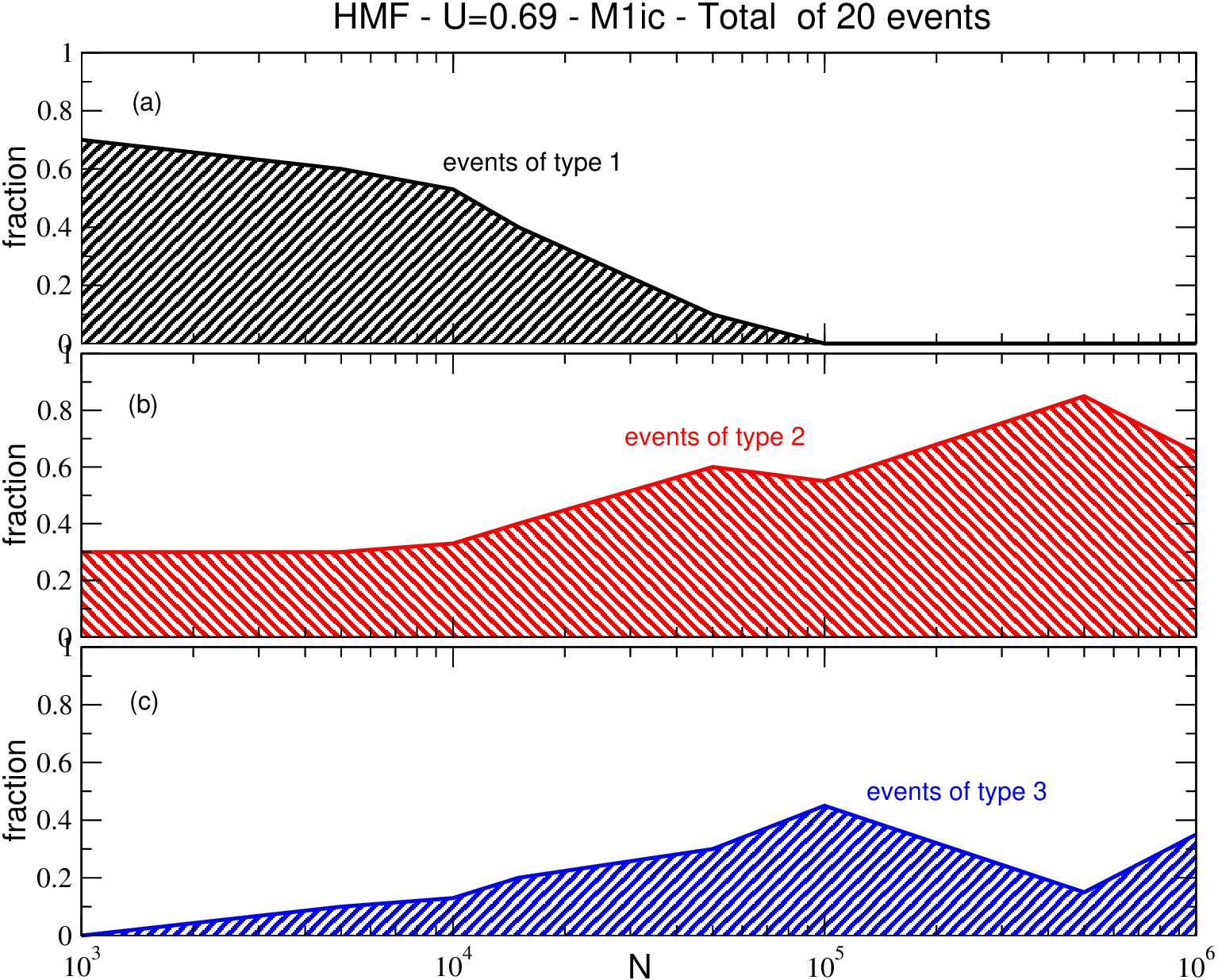}
\caption{Fraction of events of the three classes as a function of the size of the system $N$, calculated
over sets of $20$ events for each size.}
\end{figure}

In order to clarify what is the relative weight of these different kinds of events when increasing the size
of the system, we report in Fig. 3 the fraction of events of the three classes as a function
of $N$ (calculated over sets of $20$ events for each size). As one could expect, the fraction 
of \textit{type 1} events (a), initially around $0.8$, decreases with the size of the system and becomes null 
above $N=10^5$ while, on the contrary, the fraction of \textit{type 3} events (c) increases with $N$ and,
above $N=10^5$, starts to oscillate around $0.3$. On the other hand, the fraction of \textit{type 2} events (b),
those having  the double QSS plateau, stays constant around $0.3$ up to $N\sim10^4$, then rapidly increases 
and stabilizes around $0.7$ above $N\sim5\cdot10^5$: this is quite surprising because, for great values
of $N$, we know that the ensemble averaged temperature should tend to the stable value 
$T_{N\rightarrow\infty}\sim0.38$, thus one could expect that, increasing the size of the system, all the events  should belong to the \textit{type 3} class. 
\\
The non vanishing fraction of \textit{type 2} events for any value of the studied $N$ evidently indicates
that the \textit{ensemble average} of temperature, which yields single QSS plateaux at 
$T_{N\rightarrow\infty}\sim0.38$ \cite{hmf2,pluchino,epn}, will not coincide with the \textit{time average} 
over \textit{type 2} events, due to their double QSS plateau. 
Such a result is in perfect agreement with those of \cite{epl}, where it has been shown that, due to the non-ergodicity 
of the QSS regime, the inequivalence between ensemble averages and time agerages holds for velocities Pdfs, 
and q-Gaussian attractors appear instead of the usual Gaussian ones predicted by the Central Limit Theorem
when ergodicity applies. In this respect, the double plateau of the \textit{type 2} events seems to suggest
the existence of two different attractors in the QSS regime, corresponding to the two different
temperature values of the plateau. 
\\
At the same time, in order to answer to  
the criticism mentioned at the end of the previous section \cite{yama,bouchet,moyano,ruffo},
the presented results also show that, even for $N\rightarrow\infty$,
anomalous diffusion persists at least in the first part (that one  with the higher temperature) of the QSS 
plateau in \textit{type 2} events, thus it does not appear to  simply  be a finite size effect. The eventual convergence 
to  normal diffusion is very slow  and in most real  physical cases, where both time and size are finite,  it  is not assured. On the other hand, the fact of having anomalous diffusion appears to  contradict  what found in Ref. \cite{adiff1,adiff2}, according to Ref. \cite{ruffo}. However, in this  latter paper,  stationarity of the correlation function was assumed and this is in general not true, since ageing has been found in the QSS  regime, see for examples Refs. \cite{adiff3,adiff4}.
Finally, the  different  distributions plotted in Fig. 3 as a function of N  indicate a sensible dependence on N and on the type of the event, a fact  which was previously not clearly observed  for small  sizes. In this respect it  is interesting  to notice
that  the  minimum sometimes observed in the last part of the QSS temperature  plateau, when averaging over many events \cite{zanette}, could be now likely explained as the result of a mixture  of events of the three types. 

\section{Conclusions}

We have shown that three different classes of events exist in the metastable QSS regime of the HMF model,
with a relative frequency which depends on the size of the system. 
Each class presents different
details in the diffusive behavior and  some anomalies, like superdiffusion and double QSS temperature
plateau, persist also increasing the size of the system.
These are only preliminary results: in a future study we will calculate more accurately the fraction of events
of the three classes (using larger sets of events) and we will try also to estimate the crossover time between the two QSS
plateaux in the \textit{type 2} events as a function of the system size.
However, we think  that the present work  contributes to clarify some controversial points about
the anomalous behavior observed in the QSS regime of the HMF model.

\section{Acknowledgements}

We would like to thank Stefano Ruffo and Constantino Tsallis  for interesting  discussions. The numerical calculations
here presented were done within the TRIGRID
project. The authors acknowledge financial support from the PRIN05-MIUR project \textit{"Dynamics and Thermodynamics
of Systems with Long-Range Interactions".}



\end{document}